# Colossal Switchable Photocurrents in Topological Janus Transition-Metal Dichalcogenides

Haowei Xu,[1] Hua Wang,[1] Jian Zhou,[1] Yunfan Guo,[2] Jing Kong[2] and Ju Li[1,3 †]

[1] Department of Nuclear Science and Engineering, Massachusetts Institute of Technology, Cambridge, Massachusetts 02139, USA

[2] Department of Electrical Engineering and Computer Science, Massachusetts Institute of Technology, Cambridge, Massachusetts 02139, USA

[3] Department of Materials Science and Engineering, Massachusetts Institute of Technology, Cambridge, Massachusetts 02139, USA

## Abstract

Nonlinear optical properties such as bulk photovoltaic effects possess great potentials in energy harvesting, photodetection, rectification, etc. To enable efficient light-current conversion, materials with strong photo-responsivity are highly desirable. In this work, we predict that monolayer Janus transition metal dichalcogenides (JTMDs) in the 1T' phase possess colossal nonlinear photoconductivity owing to their topological band crossing, strong inversion breaking, and small electronic bandgap. 1T' JTMDs have inverted bandgaps on the order of 10 meV and are exceptionally responsive to light in the terahertz (THz) range. By first-principles calculations, we reveal that 1T' JTMDs possess shift current (SC) conductivity as large as 2300 nm · μA/V$^2$, equivalent to a photo-responsivity of 2800 mA/W. The circular current (CC) conductivity of 1T' JTMDs is as large as $\sim 10^4$ nm · μA/V$^2$. These remarkable photo-responsivities indicate that the 1T' JTMDs can serve as efficient photodetectors in the THz range. We also find that external stimuli such as the in-plane strain and out-of-plane electric field can induce topological phase transitions in 1T' JTMDs, and that the SC can abruptly flip their directions with topological transition. The abrupt change of the nonlinear photocurrent can be used to characterize the topological transition, that has potential use in 2D optomechanics and nonlinear optoelectronics.

[†] correspondence to: liju@mit.edu

With the development of strong light sources, nonlinear optical (NLO) materials have the potential to boost optoelectronic behaviors beyond the standard linear optical effects. In principle, any materials lacking inversion symmetry can be candidates for NLO applications. Recently, the generation of nonlinear direct photocurrent upon light illumination has evoked particular interest. This is known as bulk photovoltaic effect (BPVE). The photocurrent under linearly polarized light, or the shift current (SC), has been theoretically predicted and experimentally observed in materials such as multiferroic perovskites[1–6] and monolayer monochalcogenides[7–9]. The BPVE is a promising alternative source of photocurrent for energy harvesting. Compared with the conventional solar cells based on pn junctions, BPVE is not constraint by the Shockley-Queisser limit[10] and can produce open-circuit voltage above the bandgap[3]. Besides SC, the circular photogalvanic effect[11–15] (CPGE) that generates circular current (CC) (aka injection current) under circularly polarized light is yet another nonlinear photocurrent effect. In time-reversal invariant systems, SC is the response under linearly polarized light, while CC is the response under circularly polarized light, and the direction of CC can be effectively controlled by the handedness of circularly polarized light.

Besides energy harvesting, the nonlinear photocurrent effects can also be utilized for photodetection, especially in the mid-infrared (MIR) to terahertz (THz) regions, where efficient photodetectors are highly desirable. Compared with traditional infrared detectors such as MCT (HgCdTe) detector, photodetectors based on nonlinear photocurrent do not require biasing, hence the dark current can be minimized, which is advantageous especially at elevated temperatures. Topological materials may be promising candidates for the NLO photodetection. For example, Weyl semimetals (WSMs) have singular Berry curvature around the Weyl nodes, leading to strong linear and nonlinear optical responses[16–21]. Recently, the unoptimized third-order photo-responsivity of WSM $TaIrTe_4$ is reported to be 130.2 mA/W under 4 μm illumination at room temperature[20], comparable with that of state-of-art MCT detectors (600 mA/W) operating at low temperature[22,23]. Meanwhile, many other WSMs are predicted to have even larger second-order photo-responsivity[21].

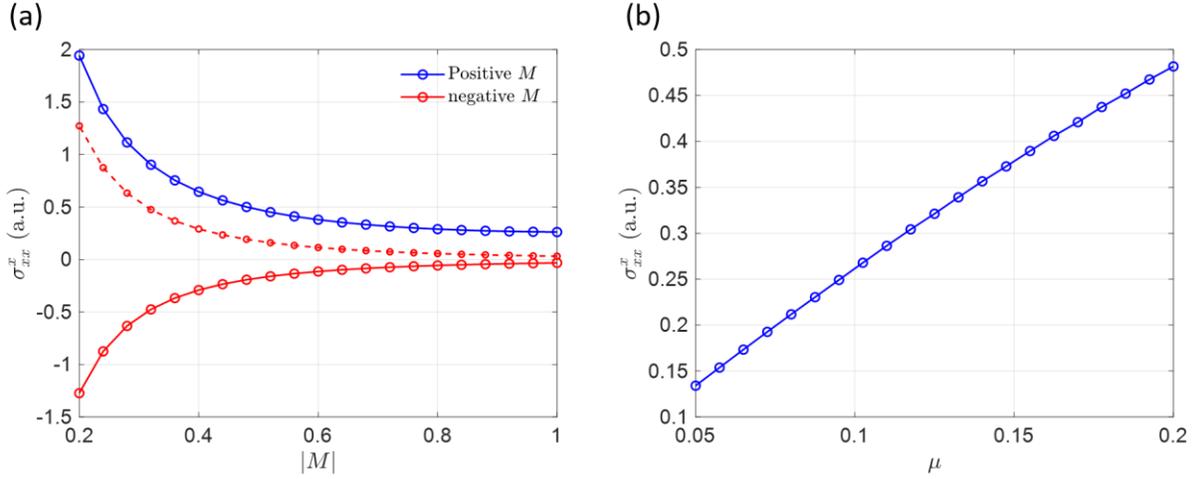

**Figure 1.** The shift current conductivity $\sigma_{xx}^{x}$ for the model Hamiltonian $H(\mathbf{k}) = \mathbf{d}(\mathbf{k}) \cdot \boldsymbol{\sigma} + \mu\sigma_x$, where $\mathbf{d}(k) = [Ak_x, Ak_y, M - B(k_x^2 + k_y^2)]$ and $\boldsymbol{\sigma} = [\sigma_x, \sigma_y, \sigma_z]$ are Pauli matrices. (a) $M$ is varied. For positive and negative $M$ with the same absolute value, $|\sigma_{xx}^{x}|$ is larger when $M$ is positive due to band inversion. $A = 2, B = 1, \mu = 0.1$ is used. Blue and red solid curves are $\sigma_{xx}^{x}$ for $M > 0$ and $M < 0$, respectively. For better visibility, the red solid curve is flipped (red dashed curve, with an extra minus sign) to be compared with the blue curve. (b) $\mu$ is varied. $\sigma_{xx}^{x}$ scales approximately linearly with $\mu$, which determines the strength of inversion symmetry breaking. $A = 2$, $B = 1$, $M = 1$ is used.

Compared to WSMs in 3D, which have vanishing bandgap and may lead to the overheating problem under strong light, 2D topological insulators (TIs) with finite bandgap on the order of $0.01 - 0.1$ eV (within the MIR/THz range) may be a better choice, thanks to their good optical accessibility and easy band dispersion manipulation. (As a matter of nomenclature, despite small bandgap values $\sim k_B T_{\text{room}}$, we still call these materials "insulators" due to the literature convention of topological insulators.) Due to the band inversion, TIs also have augmented Berry connections near the bandgap, which could enhance their optical responses[24,25]. In order better illustrate this effect, we first adopt a generic and minimal two-band model that can describe the band-inversion process[25] $H_0(\mathbf{k}) = \mathbf{d}(\mathbf{k}) \cdot \boldsymbol{\sigma}$, where $\boldsymbol{\sigma} = [\sigma_x, \sigma_y, \sigma_z]$ are Pauli matrices, and $\mathbf{d}(k) = [Ak_x, Ak_y, M - B(k_x^2 + k_y^2)]$, with $M, A$ and $B$ as model parameters. Without loss of generality, we assume $A, B > 0$ here. When $M > 0$, the mass term $M - B(k_x^2 + k_y^2)$ is positive when $k_x^2 + k_y^2$ is small, and becomes negative when $k_x^2 + k_y^2$ is large. Hence there is a band inversion process. On the other hand, when $M < 0$, the mass term $M - B(k_x^2 + k_y^2)$ is always negative, and there is no band inversion. In order to obtain finite nonlinear optical current responses, the inversion symmetry needs to be broken. Hence, we add an inversion symmetry breaking term $H_{\text{IB}} = \mu\sigma_x$ in the model Hamiltonian, where $\mu$ is a tunable parameter that controls the strength of the inversion asymmetry. Finally, $p_x = A\sigma_x$ and $p_y = A\sigma_y$ are the momentum operators.

In Ref. [25], it was demonstrated that band inversion ($M > 0$) would enhance the interband transition matrix $\langle c|r|v \rangle$ (Figure 1 therein), where $|c\rangle$ and $|v\rangle$ are the wavefunction of the conduction and valence band, $r$ is the position operator. This is due to the orbital character mixture when band inversion occurs (e.g., both $p$ and $d$ orbital components are mixed in the valence and conduction bands of 1T' TMD monolayers[26] due to band inversion). Note that $|\langle c|r|v \rangle|$ determines the response strength of the shift and circular currents, thus it should be expected that the band inversion would boost the nonlinear photocurrent responses.

Then we can calculate the shift current response function $\sigma_{xx}^x$ for the model Hamiltonian above. We first set $A = 2, B = 1, \mu = 0.1$, and vary $M$. The results are shown in Figure 1(a). One can see that when $M$ is positive (with band inversion, blue curve), $|\sigma_{xx}^x|$ is ~3 times larger than that when $M$ is negative (no band inversion, red curves) with the same absolute value $|M|$. This clearly shows that band inversion can boost the nonlinear photocurrent responses for low frequencies near the bandgap. Besides, one can see that for positive and negative $M$, $\sigma_{xx}^x$ has different signs, indicating that the photocurrents flow in opposite directions[24]. Another remarkable feature is that, when $|M|$ becomes smaller, the magnitude of the photoconductivity would increase, and there is a rough scaling relation $|\sigma_{xx}^x| \sim 1/|M|$. Note that in the current model, $|M|$ measures the bandgap ($E_g \sim 2|M|$). Hence, we suggest that small bandgaps would also boost the nonlinear photoconductivity. We would like to note again that it is the band inversion, rather than the topological nature, that boosts the nonlinear photocurrent. Materials with band inversion can be topologically trivial. Furthermore, the magnitude of the photocurrent response is also dependent on the strength of inversion asymmetry. To elucidate this effect, we fix $A = 2$, $B = 1$, $M = 1$ and vary $\mu$. The results are shown in Figure 1(b). One can see that $\sigma_{xx}^x$ scales approximately linearly with $\mu$.

The model above suggests that materials with 1) band inversion, 2) strong inversion asymmetry, and 3) small bandgaps may well have large nonlinear photoconductivity. Guided by these principles, we predict that monolayers of Janus transition metal dichalcogenides (JTMDs, denoted as MXY, $M = Mo, W$, $X, Y = S, Se, Te$) in their 1T' phase possess colossal nonlinear photocurrent conductivity with first-principles calculations. Being TIs [26], 1T' JTMDs enjoy enhanced optical responses due to the band inversion, and the maximum SC conductivity is found to be around 2300 nm · μA/V² in THz range. Such colossal SC conductivity is also about ten-folds larger than that of many WSMs reported by far[21] and other non-centrosymmetric 2D materials, such as 2H TMDs[27] and monochalcogenides[7–9]. The CC conductivity of 1T' JTMDs is also extremely large. The peak value of the CC conductivity is around $8.5 \times 10^3$ nm · μA/V² with a carrier lifetime of 0.2 ps. Owing to the small bandgap (∼ 10 meV), the SC conductivity peaks lie within the THz region, and quickly decay with increasing light frequency. The inert responsivity to light with higher frequencies renders 1T' JTMDs selective photodetectors in the THz range. Furthermore, we find that the band topology and Rashba splitting of valence and conduction bands of 1T' JTMDs can be effectively switched/tuned by small external stimuli such as in-plane strain or out-of-plane electric field. We show that such topological phase transition could lead to a sign change of the SC conductivity (and the SC direction) while maintaining its large order of magnitude. Such a colossal and switchable photocurrent may find applications in 2D optomechanics, nonlinear optoelectronics, etc. Besides, by tuning the Fermi level, the photoconductivity can be further enhanced. Besides nonlinear photoconductivity, other nonlinear optical effects, such as second-order harmonic generation, are boosted in JTMDs as well.

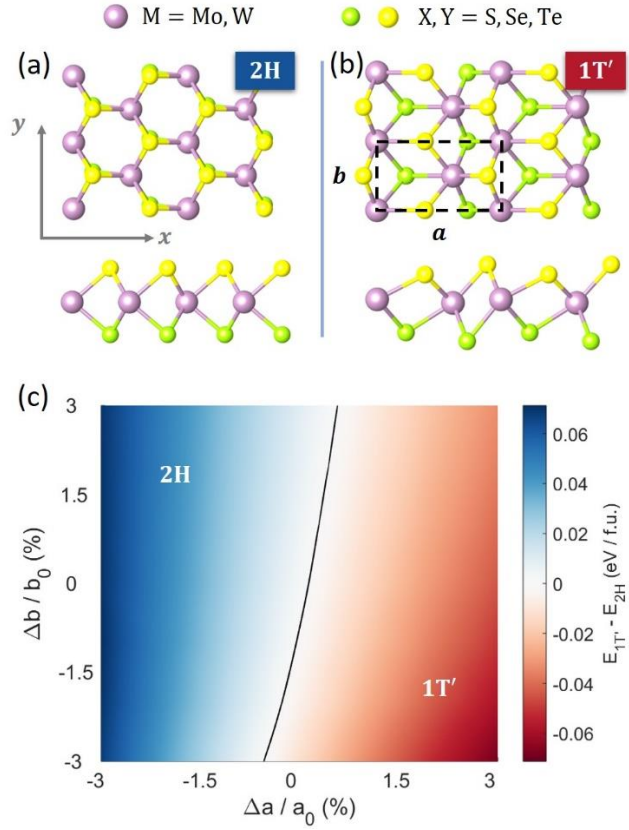

**Figure 2. (a, b)** Atomic structure of 2H and 1T' phase of JTMD. The black box in (b) shows the unit cell of 1T' phase. The top and bottom chalcogens are shifted a bit for better visibility. **(c)** The 2H-1T' phase diagram of WSeTe. $a_0$ and $b_0$ are the fully relaxed lattice constants of the 1T' phase.

The monolayer JTMDs are composed of three atomic layers: a middle layer of transition metals is sandwiched by two side layers of different chalcogen atoms (Figure 2). Inherited from pristine TMDs (PTMDs), JTMDs also have different crystalline phase structures. Among them, the 2H and 1T′ are two (meta-)stable structures. The 2H phase JTMDs (space group P3m1, Figure 2a) have a quasi-Bernal (ABA′) stacking pattern with three-fold in-plane rotational symmetry and have been successfully fabricated recently[28–30]. On the other hand, the 1T′ phase (space group Pm, Figure 2b) has an ABC stacking pattern, and the in-plane rotational symmetry is broken by a Peierls distortion along the $x$-axis. With fully relaxed lattice constants, all six MXY have lower energy in 2H phase than in 1T′ phase[31], with a small energy difference ($\lesssim 0.1$ eV/f.u.). Similar to the PTMDs, the relative stability of these two phases can be effectively tuned by strain (see SM). For example, we plot the phase diagram of WSeTe in Figure 2c, which clearly suggests a tensile strain of less than 1% along the $x$-axis can render the 1T′ phase more stable. Also, the energy barriers between 2H and 1T' phases are high ($\gtrsim 1$ eV/f.u.), thus 1T' JTMDs are fairly stable even in the strain-free state.

The JTMDs inherit many salient properties of PTMDs. As the top-bottom chalcogen layers break the inversion (mirror) symmetry of the 1T′ (2H) phase, JTMDs possess extra properties apart from those of PTMDs, such as larger Rashba spin splitting[32,33], more efficient charge separation[34], etc. The 1T′ PTMDs are $Z_2$ topological insulators[26] with small bandgaps on the order of 10 meV, indicating a strong optoelectronic coupling in the THz range, because both inverted band structure and small bandgaps would enhance the interband transitions. However, due to the centrosymmetry, the second-order NLO effects are forbidden for 1T' PTMDs. On the contrary, 1T′ JTMDs are inherently non-centrosymmetric owing to the two different chalcogen layers, and the giant second-order NLO effects can be unleashed.

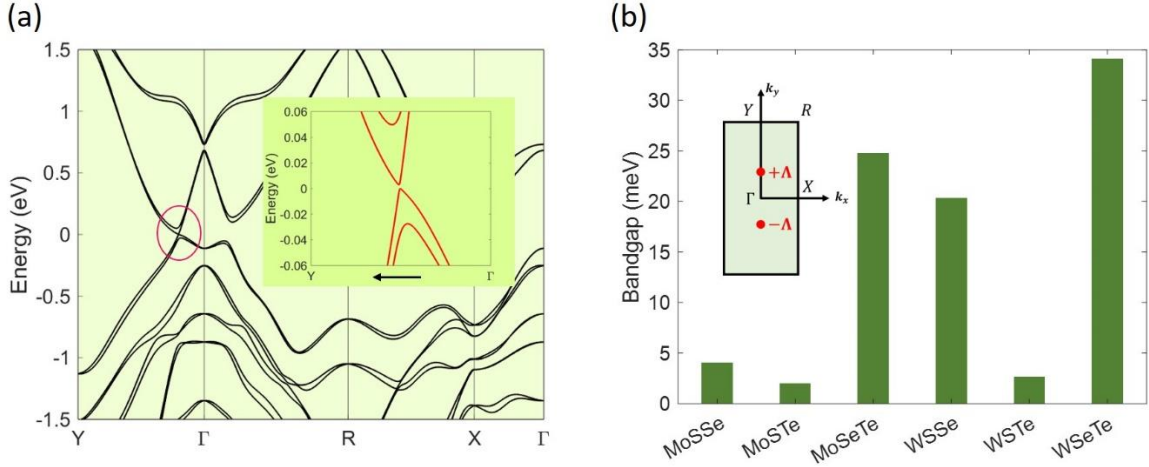

**Figure 3 (a)** Band structure of 1T' MoSSe, the energy offset is moved to the top of the valence band. Inset: band structure around the fundamental bandgap $\Lambda$ (along $\Gamma$-Y). It is clear that the strong Rashba splitting breaks the two-fold degeneracy. **(b)** Fundamental bandgaps of all six MXY. Inset: first BZ of 1T' JTMDs.

Considering MoSSe as an example, we show the electronic properties of 1T′ JTMDs. The band structure of MoSSe is shown in Figure 3a. Like PTMDs[26], the metal $d$-orbitals and chalcogen $p$-orbitals are inverted around the $\Gamma$ point, and the inverted bandgap is around 0.8 eV. The fundamental bandgaps are along the $\Gamma$-Y line ($\pm\Lambda$ point, inset of Figure 3b) with a magnitude of $E_g \approx 4$ meV. We find that the fundamental bandgaps of all six 1T′ JTMDs lie in the range of 1 ~ 50 meV, corresponding to the THz range (Figure 3b). Interestingly, despite the band inversion around the $\Gamma$ point, not all 1T′ JTMD are topologically nontrivial. With fully relaxed atomic structures, MSSe and MSeTe have $Z_2 = 0$ while MSTe have $Z_2 = 1$ (M = W, Mo. $Z_2 = 0$ and 1 indicate trivial and nontrivial band topology, respectively). This is because the large Rashba splitting from the inversion symmetry-breaking could change the band topology by remixing the wavefunctions around the $\pm\Lambda$ point. As we will show later, both in-plane strain and out-of-plane electric field can induce a topological phase transition by closing and reopening the fundamental bandgap[35–38]. Regardless of the band topology ($Z_2$ number), the band inversion around $\Gamma$ point gives rise to a strong wavefunction mixing between valence bands (VB) and conduction bands (CB)[25], which could significantly boost the linear and nonlinear responses.

In materials without inversion symmetry $\mathcal{P}$, NLO direct currents (dc) can be generated upon photo-illumination. This current can be divided into two parts, the shift current $j_{SC}$ and the circular current $j_{CC}$

$$j_{SC}^c = 2\sigma_{ab}^c(0;\omega,-\omega)E^a(\omega)E^b(-\omega) \qquad (1)$$

$$\frac{dj_{CC}^c}{dt} = 2\eta_{ab}^c(0;\omega,-\omega)E^a(\omega)E^b(-\omega)$$

where $a, b, c$ are Cartesian indices and $E(\omega)$ is the Fourier component of the optical electric field at angular frequency $\omega$. Eq. (1) indicates that when the optical electric field has both $a$ and $b$ components ($a$ and $b$ can be the same), there will be a direct current along the $c$-th direction when $\sigma_{ab}^c / \eta_{ab}^c$ is non-zero. In materials with time-reversal symmetry $\mathcal{T}$, the response functions within the independent particle approximation in clean, cold semiconductors are[39]

$$\sigma_{ab}^c(0;\omega,-\omega) = -\frac{e^3}{2\hbar^2}\int\frac{d\mathbf{k}}{(2\pi)^3}\sum_{n,m} f_{nm}\frac{r_{mn}^a r_{nm;c}^b + r_{mn}^b r_{nm;c}^a}{\omega_{mn} - \omega - i/\tau} \qquad (2)$$

$$\eta_{ab}^c(0;\omega,-\omega) = -\frac{ie^3}{2\hbar^2}\int\frac{d\mathbf{k}}{(2\pi)^3}\sum_{n,m} f_{nm}\frac{\Delta_{mn}^c[r_{mn}^a, r_{nm}^b]}{\omega_{mn} - \omega - i/\tau}$$

Here all dependencies on $\mathbf{k}$ are omitted. $\tau$ is the carrier lifetime. $m, n$ are band indices, while $f_{mn} \equiv f_m - f_n$, $\omega_{mn} = \omega_m - \omega_n$, $\Delta_{mn} = v_{mm} - v_{nn}$ are the differences in occupation number, energy, and band velocity between bands $n$ and $m$, respectively. $r_{mn} = i\langle m|\nabla_k|n\rangle$ is the Berry connection, $[r_{mn}^a, r_{nm}^b] = r_{mn}^a r_{nm}^b - r_{mn}^b r_{nm}^a$ is the Berry curvature, while $r_{mn;c}$ is the generalized gauge covariant derivative of $r_{mn}$, defined as $r_{mn;c}^b = \frac{dr_{mn}^b}{dk_c} - i(\xi_{mm}^c - \xi_{nn}^c)r_{mn}^b$, where $\xi_{mm} = i\langle m|\nabla_k|m\rangle$ is the intraband Berry connection. Even though the interband Berry connection is gauge dependent, its generalized derivative $r_{mn;c}^b$ is gauge invariant. Eq. (2) is slightly different from those in Ref. [39] by explicitly including $\tau$-dependence.

When the carrier lifetime satisfies $\tau \gg \hbar/E_g$, the $i/\tau$ term in the denominator of Eq. (2) can be neglected. In this case, $\sigma_{ab}^c(0;\omega,-\omega)$ is purely real, while $\eta_{ab}^c(0;\omega,-\omega)$ is purely imaginary. Considering that the direct current should be a real quantity, $E^a$ and $E^b$ should have 0 ($\frac{\pi}{2}$) phase difference to yield non-vanishing SC (CC), which indicates that SC and CC are responses under linearly and circularly polarized light, respectively. Another noteworthy feature is that, upon light illumination, $j_{CC}$ grows with time at the initial stage, and the saturated static CC should be $j_{CC} = \tau\eta_{ab}^c E^a E^b$, with $\tau$ as the carrier lifetime. Therefore $\tau\eta$ can be regarded as the effective CC photoconductivity.

Another formula describing the nonlinear photocurrents can be obtained from quadratic Kubo response theory[40,41], and reads

$$j^c = \frac{e^3}{2\omega^2 \hbar^2} \text{Re} \left\{ \sum_{l,m,n}^{\Omega=\pm\omega} \int \frac{d\bm{k}}{(2\pi)^3} f_{ln} \frac{v_{nl}^a}{\left(\omega_{nl} - \Omega - \frac{i}{\tau}\right)} \left[ \frac{v_{lm}^b v_{mn}^c}{(\omega_{nm} - i/\tau)} - \frac{v_{lm}^c v_{mn}^b}{(\omega_{ml} - i/\tau)} \right] E_a(\Omega) E_b(-\Omega) \right\} \quad (3)$$

Where $\omega$ is the frequency of the light. Here $v_{nl} \equiv \langle n|\hat{v}|l\rangle$ is the velocity matrix element. Eq. (2) uses the length gauge, while Eq. (3) uses the velocity gauge. It can be shown (SM) that in the presence of time-reversal symmetry $\mathcal{T}$, Eq. (3) is generally equivalent to Eqs. (1, 2), and the real and imaginary parts of Eq. (3) correspond to the SC and CC, respectively. Compared with Eqs. (1,2), Eq. (3) is more general. For example, it can be used to calculate photocurrents in magnetic materials where $\mathcal{T}$ is broken[42,43]. However, numerically Eq. (3) can experience convergence problems at small $\omega$. Therefore, Eqs. (1, 2) are adopted for computations in this work, which do not involve magnetism. The consistency between these two methods are well tested. In practice, the Brillouin zone (BZ) integration is carried out by $\bm{k}$-mesh sampling with $\sigma_{3D} = \int \frac{d\bm{k}}{(2\pi)^3} I(\bm{k}) = \frac{1}{V} \sum_{\bm{k}} w_{\bm{k}} I(\bm{k})$, where $V$ is the volume of the unit cell, $w_{\bm{k}}$ is weight factor, and $I(\bm{k})$ is the integrand. However, for 2D materials, the definition of volume $V$ is ambiguous, because the thickness of 2D materials is ill-defined[44]. Thus, we replace volume $V$ with the area $S$, and define $\sigma_{2D} = \frac{1}{S} \sum_{\bm{k}} w_{\bm{k}} I(\bm{k})$. Note that all ingredients $S, w_{\bm{k}}$ and $I(\bm{k})$ are well-defined and can be directly obtained from numerical computations, hence $\sigma_{2D}$ is unambiguous for 2D materials. As a result, in this work we mainly show $\sigma_{2D}$. The 2D and 3D conductivities satisfy $\sigma_{2D} = L_{eff} \sigma_{3D}$, where $L_{eff}$ should be the effective thickness of the material (not the thickness of the computational cell, which includes the thickness of the vacuum layer). $L_{eff}$ has no standard definitions and is usually set as the inter-layer distance when the monolayers are stacked along $z$ direction. We use an effective thickness of $L_{eff} = 6$ Å for JTMDs when $\sigma_{3D}$ is required for e.g., the comparison with other materials. Unless explicitly stated, the carrier lifetime is set as $\tau = 0.2$ ps, which should be a conservative value considering that the carrier lifetimes of 2H TMDs are more than 1 ps at room temperature[45,46].

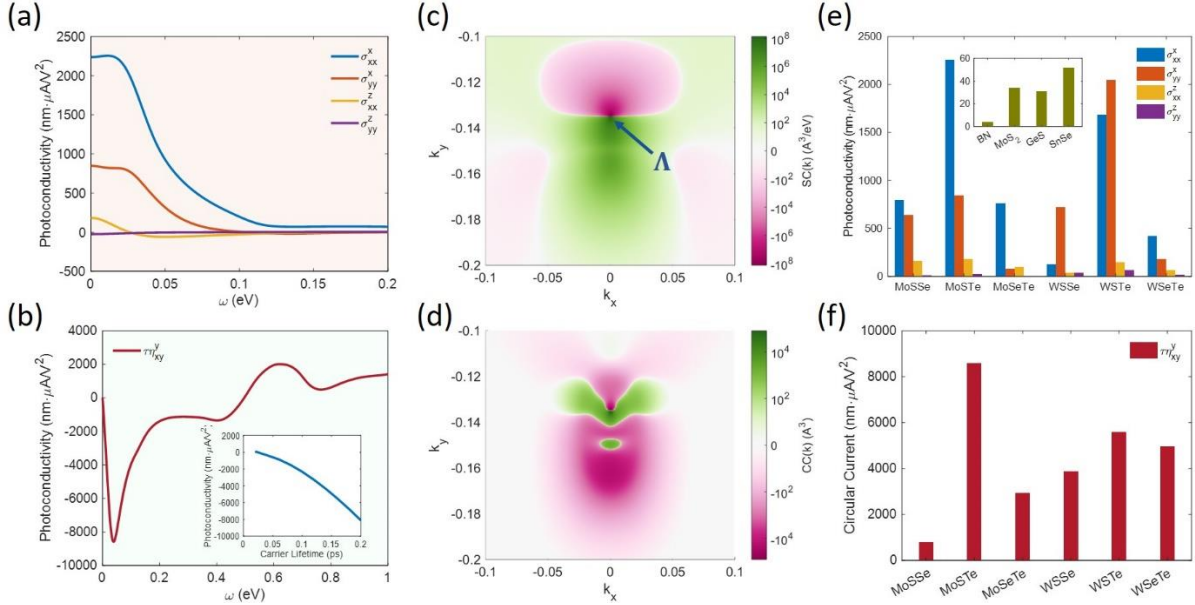

**Figure 4 (a)** The SC conductivity $\sigma_{ab}^c(0;\omega,-\omega)$ of MoSTe. **(b)** The CC conductivity $\tau\eta_{ab}^c(0;\omega,-\omega)$ of MoSTe. Inset: the first peak CC conductivity as a function of carrier lifetime $\tau$. **(c, d)** the $k$-specified contribution to $\sigma_{xx}^x$ and $\eta_{xy}^y$. (c) shows $SC(\mathbf{k})$ at $\omega = 10\ meV$ while (d) shows $CC(\mathbf{k})$ at $\omega = 50\ meV$. The colormaps are in logarithmic scale. $k_x$ and $k_y$ are in the unit of reciprocal lattices **(e, f)** the peak values of the SC (e) and CC (f) conductivities of six MXY. Inset of (e): the peak values of SC conductivities of several other 2D materials including hBN, 2H MoS$_2$, GeS and SnSe.

Note that 1T' JTMDs have mirror symmetry $\mathcal{M}^y$. The $y$-th components of $j$ and $E$ should be flipped under $\mathcal{M}^y$, while other components do not change. Consequently, $\mathcal{M}^y$ enforces $\sigma_{ab}^c/\eta_{ab}^c$ to be zero when there is an odd number of $y$ in $[a,b,c]$, such as $\sigma_{xx}^y$. The different nonzero SC conductivities of 1T' MoSTe are plotted in Figure 4a. We observe that both in-plane polarizations $\sigma_{xx}^x$ and $\sigma_{yy}^x$ have striking magnitudes of $\gtrsim 10^3$ nm · µA/V$^2$ in the THz range ($\omega <$ 10 THz $\approx$ 41 meV). The peak values of $\sigma_{xx}^x$ and $\sigma_{yy}^x$ are around 2300 nm · µA/V$^2$ and 850 nm · µA/V$^2$ respectively, more than ten-fold larger than those of other non-centrosymmetric 2D materials, such as hexagonal BN (hBN), 2H MoS$_2$, GeS, and SnSe, which are on the order of 10 $\sim$ 100 nm · µA/V$^2$ (inset of Figure 4e). When light with intensity 10 mW/cm$^2$ is shining on single layer MoSTe with 1 cm × 1 cm dimension, the photocurrent generated is on the order of 1 nA. Note that the nonlinear photocurrent can be boosted by 1) focusing the light beam. For example, when the light with the same total power is focused onto a 0.01 cm$^2$ spot size, the electric field is enhanced $10^1\times$, the photocurrent density would be $10^2\times$ and the total photocurrent would be 100 nA; 2) stacking single layer detectors to increase the cross-section. Notably, the SC conductivities quickly decay for $\omega \gtrsim 0.1$ eV, indicating that 1T' MoSTe is relatively insensitive to light beyond the THz range, which can be advantageous when selective photodetectors in the THz range are desired. Besides, an in-plane electric field can induce a large photocurrent in the out-of-plane direction: $\sigma_{xx}^z$ and $\sigma_{yy}^z$ have peak values of 180 and 25 μA/V$^2$, respectively. Such an out-of-plane current can be measured if transparent electrodes like graphene are attached directly above and below the MoSTe monolayer.

To understand the origin of such giant photoconductivity, the $\mathbf{k}$-specific contribution to the total SC conductivity, $SC(\mathbf{k}) \equiv \mathrm{Re}\left\{\sum_{n,m} f_{nm} \frac{r_{mn}^a r_{nm;c}^b + r_{mn}^b r_{nm;c}^a}{\hbar(\omega_{mn}-\omega-i/\tau)}\right\}$ at $\omega = 10$ meV is shown in Figure 4c. We can see that around the fundamental bandgap Λ, $SC(\mathbf{k})$ has a peak amplitude of about $\pm 10^8$ Å$^3$/eV. Away from Λ, $SC(\mathbf{k})$ rapidly decays. This phenomenon is consistent with the argument that the inverted band structure would lead to enhanced Berry-connection magnitudes. We also calculate the SC conductivity for the other five 1T′ JTMDs, and their peak values are shown in Figure 4e. All of six 1T′ JTMDs possess colossal photovoltaic effect and the peak values of $\sigma_{xx}^x$ and $\sigma_{yy}^x$ are on the order of $10^3$ nm·μA/V$^2$. Generally, MSTe exhibits stronger BPVE than MSSe and MSeTe. This is due to a larger out-of-plane asymmetry in the MSTe system. The electron affinity of S, Se, and Te atoms are 2.08, 2.02, 1.97 eV, respectively. Consequently, the out-of-plane asymmetry should be more significant in MSTe, leading to stronger BPV. This point is also verified by the out-of-plane electric dipole $P_z$. We find that $P_z$ of MSTe is around 0.15 $e$·Å per unit cell, while for MSSe and MSeTe, $P_z$ is only around 0.07 ~ 0.08 $e$·Å per unit cell.

The CC conductivities are plotted in the lower panels of Figure 4 (Fig. 4b,d,f). With in-plane polarization, the only non-vanishing element of CC tensor is $\eta_{xy}^y$, based on the symmetry analysis above. $\tau\eta_{xy}^y$ of MoSTe has a peak value of $8.5 \times 10^3$ nm·μA/V$^2$ around $\omega \approx 50$ meV (Figure 4b). Since $\tau\eta_{xy}^y$ is sensitively dependent on the carrier lifetime, we vary $\tau$ and obtain the peak values of $\tau\eta_{xy}^y$ (inset of Figure 4b). Even with $\tau = 0.04$ ps, $\tau\eta_{xy}^y$ still has a peak value of around 400 nm·μA/V$^2$. The $\mathbf{k}$-specific contribution to the total CC conductivity, $CC(\mathbf{k}) \equiv \mathrm{Re}\left\{\sum_{n,m} f_{nm} \frac{\Delta_{mn}^c [r_{mn}^a, r_{nm}^b]}{\omega_{mn}-\omega-i/\tau}\right\}$ at $\omega = 50$ meV, is plotted in Figure 4d. Similar to SC, the major contributions also lie in the vicinity of Λ. Finally, the peak values of $\tau\eta_{xy}^y$ for all six 1T′ JTMDs are shown in Figure 4f. Similar as in SC, the CC conductivity in MSTe, which has stronger inversion asymmetry, is stronger than those in MSSe and MSeTe. Besides SC and CC, which are

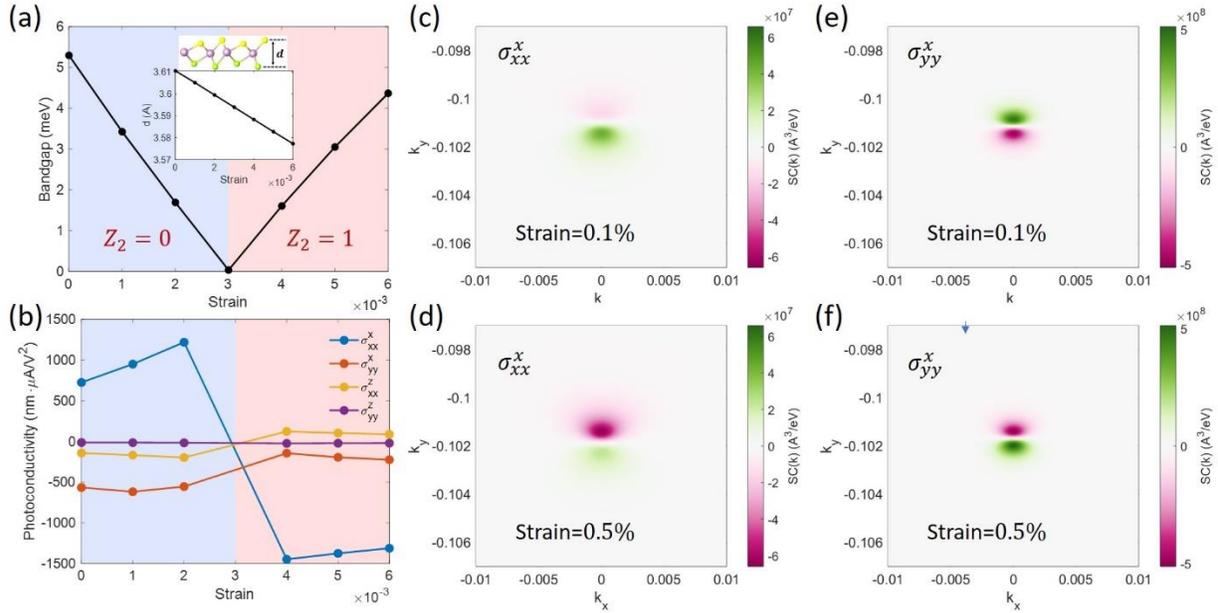

**Figure 52 (a)** The bandgap of 1T' MoSSe as a function of biaxial in-plane strain $\epsilon$. When $\epsilon < 0.3\%$, 1T' MoSSe is topologically trivial. While for $\epsilon > 0.3\%$, it is a $Z_2$ topological insulator. Inset: the atomic thickness of 1T' MoSSe as a function of $\epsilon$. The vertical chalcogen distance shrinks with tensile strain, thus altering the magnitude of the Rashba splitting. **(b)** The SC conductivities of MoSSe as a function of $\epsilon$. There are abrupt jumps when MoSSe goes through the topological transition, and $\sigma_{xx}^x$ and $\sigma_{xx}^z$ flip directions. This effect can be understood with the $k$-specified contribution to the total SC conductivity $SC(k)$ in **(c-f)**.

interband contributions to the nonlinear photocurrent, there could also be intraband contributions. For insulating materials at zero temperatures, the intraband part should be zero. But since 1T' JTMDs have small bandgaps comparable with room temperature ($k_B T_{\text{room}} \sim 26$ meV), we have also calculated the intraband contribution due to anomalous velocity at finite temperatures. The results are shown in the SM, and one can find that the intraband contributions can be on the same order as the interband contributions. As discussed above, around the $\pm \Lambda$ points, the Rashba splitting breaks the degeneracy and could close and reopen the bandgap, leading to topological phase transitions. The magnitude of the Rashba splitting could be engineered with external stimuli, such as in-plane strain, external electric field, etc. For example, with a tensile strain, the vertical chalcogen distance of 1T′ JTMD shrinks (inset of Figure 5a). The bandgap of MoSSe as a function of biaxial in-plane strain $\epsilon$ is plotted in Figure 5a, where a band closing occurs around $\epsilon = 0.3\ \%$. This band closing/reopening indicates a topological transition. For $\epsilon < 0.3\ \%$, 1T′ MoSSe has trivial band topology with $Z_2 = 0$. While with $\epsilon > 0.3\ \%$, 1T′ MoSSe becomes a $Z_2$ topological insulator. Such sensitive dependence on in-plane strain provides a convenient pathway to trigger topological phase transitions in 1T′ JTMD. An even more intriguing phenomenon arises in the SC responses. In Figure 5b we show the SC conductivity of 1T′ MoSSe as the function of $\epsilon$. All four components of $\sigma_{ab}^{c}$ undergo an abrupt jump upon the topological transition. Particularly, $\sigma_{xx}^{x}$ and $\sigma_{xx}^{z}$ flip their directions. Such an abrupt jump originates in the change in the band characteristics around Λ upon the topological transition[24]. As discussed above, the major contributions to the total SC conductivity come from $k$-points around Λ point (Figure 3b). When the bandgap is closed and reopened, the wavefunctions of the lowest CB and highest VB around Λ point undergo a substantial remixing. In ideal cases such as a two band model[24], $I_{mn;c}^{ab} = r_{mn}^{a} r_{nm;c}^{b} + r_{mn}^{b} r_{nm;c}^{a}$ would flip sign since $m$ and $n$ is interchanged and $I_{mn;c}^{ab}$ is purely imaginary. This is consistent with the results with the model Hamiltonain before. When more band contributions are incorporated, $I_{mn;c}^{ab}$ does not always flip its sign, but would still experience a drastic change. The arguments above are verified by the $\boldsymbol{k}$-specific contribution to $\sigma_{xx}^{x}$ and $\sigma_{yy}^{x}$ as shown in Figure 5(c-f), where we can see $SC(\boldsymbol{k})$ are significantly different on two sides of the topological transition. In addition to in-plane strain, an out-of-plane electric field, which also modifies the magnitude of the Rashba splitting, can trigger the topological transition and alter the SC conductivities as well (see SM). Thus, we propose that the abrupt jump of nonlinear photocurrent is a universal signature of the topological phase transition in non-centrosymmetric materials, and can be used as an online diagnostic tool. The mechanical, electrical and even optomechanical[47,48] approaches to switching the NLO responses would pave the way for efficient and ultrafast nonlinear optoelectronics.

It is also intriguing to investigate how the nonlinear photocurrents vary when the Fermi level is buried in the CB or VB by carrier doping. The SC and CC conductivities of MoSTe as the function of the Fermi level $E_F$ are shown in Figure 6a. We can see that for $E_F$ within $\pm 50$ meV ($E_F$ is set as 0 when the Fermi level is on the top of the VB), the SC and CC conductivities remain extremely large in their magnitudes. While for $E_F$ far away from the fundamental bandgap (heavily carrier doped), both SC and CC conductivities decay to zero. Here, the pure intraband nonlinear anomalous Hall current discussed by Wang *et al.* [49] is not considered. A noteworthy feature is that, when $E_F$ is slightly above (below) the bandgap, $\sigma_{yy}^{x}$ would jump to an enormously positive (negative) value, about ten times larger in amplitude than that when $E_F$ is inside the bandgap. This effect can be understood by looking at the band structure (Figure 3a) and the $\boldsymbol{k}$-specific

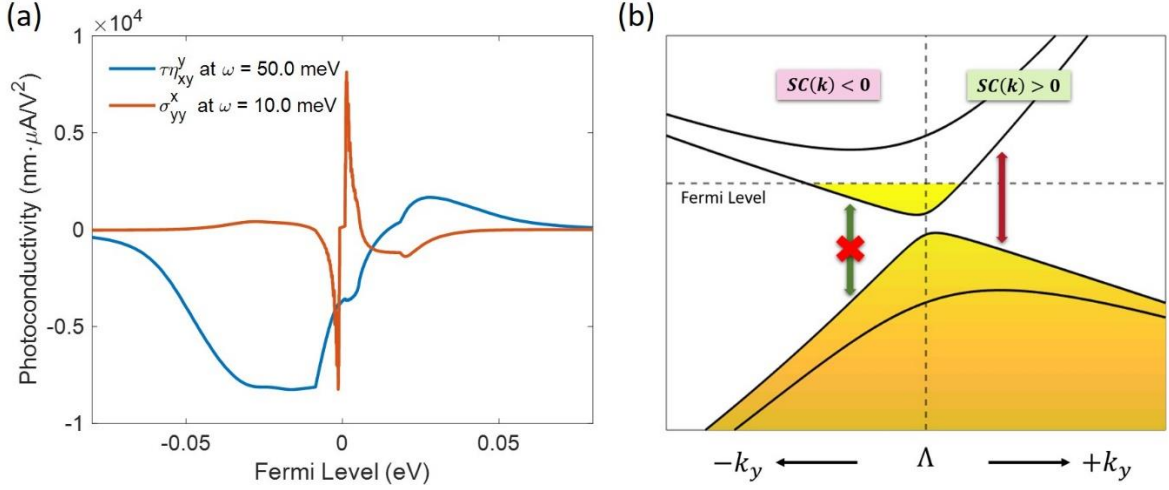

**Figure 6** (a) The CC and SC conductivities of MoSTe as a function of the Fermi level $E_F$. $E_F = 0$ denotes the Fermi level is at the valence band maximum. (b) An illustration that explains the sharp jump in the photoconductivity when $E_F$ is slightly inside the valence band or the conduction band.

contribution $SC(\mathbf{k})$ (Figure 4c). As discussed above, the major contribution to the total SC conductivity comes from $\mathbf{k}$-points close to the fundamental bandgap Λ. When the VB and CB are occupied and empty respectively, $SC(\Lambda + \delta k_y)$ and $SC(\Lambda - \delta k_y)$ ($\delta$ is a small positive parameter) have opposite values and tend to cancel each other. On the other hand, with a positive $E_F$, those CB below the Fermi level would be occupied as well, and the CB-VB transition cannot contribute to $SC(\mathbf{k})$ anymore (Figure 6b). However, a lager region on the $\Lambda - \delta k_y$ side would have occupied CB than on the $\Lambda + \delta k_y$ side. This is because the CB cone is tilted and the band velocity is smaller on the $\Lambda - \delta k_y$ side, leading to a larger partial density of states in this region. As a result, the positive $SC(\mathbf{k})$ on the $\Lambda + \delta k_y$ side would be cancelled less by the negative $SC(\mathbf{k})$ on the $\Lambda - \delta k_y$ side, leading to a larger total SC conductivity. A similar analysis could show that when $E_F$ is within the VB, the total SC would have a significant negative value. These observations indicate that the photocurrent conductivity could be further enhanced by Fermi-level tuning in materials with tilted CB and/or VB, such as type-II WSM[50]. From Figure 6a, one can see that a ~ 1 meV shift in $E_F$ can dramatically enhance $\sigma_{yy}^x$. In practice, $E_F$ can be tuned by e.g., gate voltage. Assuming a gate coupling efficiency of 0.1, then a ~10 mV gate volgate would be able to achieve the enhancement.

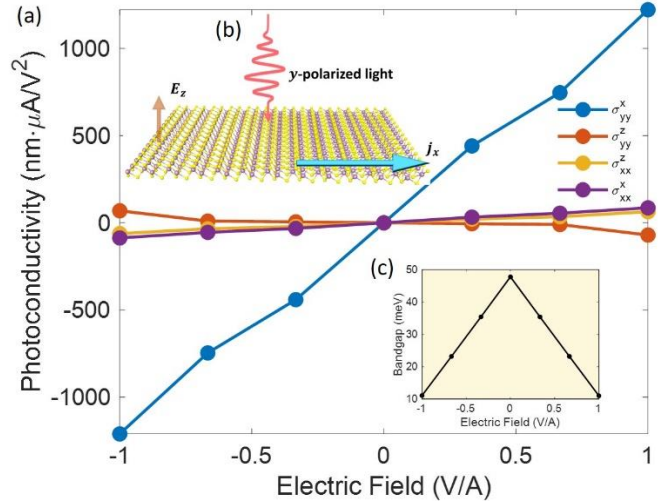

**Figure 7** (a) The SC conductivity of 1T' MoS$_2$ as a function of out-of-plane electric field. The out-of-plane electric field breaks the inversion symmetry and the photocurrent flip direction when the direction of the electric field is flipped. Inset (b): A schematic illustration of the system: A PTMD is under an out-of-plane electric field $+E_z$. A light polarized in $y$ direction shines on the PTMD and a current in the $+x$ direction $j_x$ is induced from the non-zero $\sigma_{yy}^x$. The direction of the current would flip to $-x$ if the out-of-plane electric field is flipped to $-E_z$. Inset (c): bandgap of 1T' MoS$_2$ as a function of out-of-plane electric field.

Before concluding, we would like to note that, in addition to nonlinear photocurrents, other NLO effects such as the second-harmonic generation are also colossal in 1T′ JTMDs (see SM). Besides, the inversion symmetry of 1T′ PTMDs can be broken externally by e.g., an out-of-plane electric field, resulting in nonlinear photocurrents, which can be regarded as a third-order nonlinear effect. The SC conductivity can be giant as well and can flip direction under a vertical electric field (Figure 7). Also, the SC conductivity depends approximately linearly on the electric field, which characterizes the strength of inversion asymmetry. This is consistent with results with the model Hamiltonian before, when $\mu$ plays a similar role as the electric field.

In conclusion, we reveal the colossal nonlinear photocurrent effects in 1T′ JTMDs. The photo-responsivity peaks within the THz range. As a result, the 1T′ JTMDs can be efficient and selective photodetectors in the THz range. We also investigate the topological order of 1T′ JTMDs and find that it can be conveniently switched by a small external stimulus such as in-plane strain and out-of-plane electric field. Upon the topological transitions, the photocurrents undergo an abrupt change can flip direction, which can be used as a signal of the topological transition and lead to sensitive manipulation of NLO effects. The colossal and switchable nonlinear photocurrents could find broad applications in photodetection, nonlinear optoelectronics, optomechanics, etc.

## Methods

*Ab initio* **calculations.** The first-principles calculations are based on density functional theory (DFT)[51,52], as implemented in Vienna *ab initio* simulation package (VASP)[53,54]. Generalized gradient approximation (GGA) in the form of Perdew-Burke-Ernzerhof (PBE)[55] is used to treat the exchange-correlation interactions. Core and valence electrons are treated by projector augmented wave (PAW) method[56] and a plane wave basis set with a cutoff energy of 520 eV, respectively. For the DFT calculations, the first Brillouin zone is sampled by a Γ-centered $\boldsymbol{k}$-mesh with grid density of at least $2\pi \times 0.02$ Å$^{-1}$ along each dimension. For the electric field calculations, a sawtooth-like potential along the $z$ direction is applied, with discontinuity at the middle of the vacuum layer in the simulation cell. The symmetry constraints are completely switched off in all VASP calculations to avoid incorrect handling of the electric field[57]. To further test the correctness of the bandgap-electric field relationship, we redo the calculations with Quantum Espresso[58], and the results agree well with that of VASP.

**Wannier function fittings.** The Bloch wavefunctions from DFT calculations are projected onto the maximally-localized Wannier functions (MLWF) with the Wannier90 package[59]. The MLWFs $|n\boldsymbol{R}\rangle$ are defined as

$$|n\boldsymbol{R}\rangle = \frac{1}{N}\sum_{\boldsymbol{k}} e^{-i\boldsymbol{k}\cdot\boldsymbol{R}} \sum_{m=1}^{J} U_{mn}^{\boldsymbol{k}} |m\boldsymbol{k}\rangle \quad (S1)$$

where $|m\boldsymbol{k}\rangle$ are the Bloch wavefunctions as obtained in the DFT calculations, $\boldsymbol{R}$ are Bravais lattice vectors, $J$ is the number of Wannier bands, and $U_{mn}^{\boldsymbol{k}}$ is a unitary transformation such that the Wannier functions are maximally localized. The Wannier Hamiltonian $H^W$ is constructed from the MLWFs, with

$$H^W_{nm\mathbf{R}} = \langle n0|\hat{H}|m\mathbf{R}\rangle \tag{S2}$$

Wannier Hamiltonian in the $k$ space can be obtained with a Fourier transformation

$$H^W_{nm\mathbf{k}} = \sum_{\mathbf{R}} e^{i\mathbf{k}\cdot(\mathbf{R}+\mathbf{r}_m-\mathbf{r}_n)} H^W_{nm\mathbf{R}} \tag{S3}$$

where we have included the Wannier centers $\mathbf{r}_m$ in the phase factor[60,61]. By diagonalizing $H^W_{nm\mathbf{k}}$ at each $\mathbf{k}$ point, one obtains the energy and wavefunctions $E^W_n(\mathbf{k})$ and $|n\mathbf{k}\rangle^W$.

**Band velocity, Berry connection, and sum rule.** The Wannier Hamiltonian and wavefunctions are directly applied to calculate the band velocity $v_{mn}$ with

$$v^a_{mn} = \left\langle m\left|\frac{\partial H}{\partial k_a}\right|n\right\rangle^W \tag{S4}$$

Then the interband Berry connections $r_{mn}$ can be obtained with the relation

$$r_{mn} = \frac{v_{mn}}{i\omega_{mn}} \quad (m \neq n) \tag{S5}$$

And the generalized gauge covariant derivative of $r_{mn}$ is calculated with the sum rule[8,27,60]

$$r^a_{nm;b} = \frac{i}{\omega_{nm}}\left[\frac{v^a_{nm}\Delta^b_{nm} + v^b_{nm}\Delta^a_{nm}}{\omega_{nm}} - w^{ab}_{nm} + \sum_{p\neq n,m}\left(\frac{v^a_{np}v^b_{pm}}{\omega_{pm}} - \frac{v^b_{np}v^a_{pm}}{\omega_{np}}\right)\right] \tag{S6}$$

$$(n \neq m)$$

where $\Delta_{nm} = v_{nn} - v_{mm}$, and $w^{ab}_{nm} = \frac{1}{\hbar}\left\langle n\left|\frac{\partial^2 H}{\partial k_a \partial k_b}\right|m\right\rangle^W$.

**Nonlinear photoconductivity.** After all the ingredients, $v^a_{mn}$, $r^a_{mn}$ and $r^a_{mn;b}$, are obtained from the Wannier interpolations, the nonlinear photoconductivity is calculated based on Eq. (2) in the main text. The BZ integration is sampled with a $1601 \times 3201$ $\mathbf{k}$-mesh in the first BZ. The $\mathbf{k}$-mesh convergence is tested with a denser $2251 \times 4501$ $\mathbf{k}$-mesh, and the difference is found to be negligible (SM).

**Code Availability.** The data that support the findings within this paper and the MATLAB code for calculating the shift and circular current conductivity are available from the corresponding authors upon reasonable request.

## Acknowledgments

This work was supported by an Office of Naval Research MURI through grant #N00014-17-1-2661. Y. G. and J. K. acknowledge the support from U.S. Department of Energy (DOE), Office of Science, Basic Energy Sciences (BES) under Award DE-SC0020042.

## Author Contributions

J.L. and H.X. conceived the idea and designed the project. H.X. performed the *ab initio* calculations. H.X., H.W., J.Z., and Y.G. analyzed the data. J.L. and J.K. supervised the project. All authors wrote the paper and contributed to the discussions of the results.

## Competing Interests

The authors declare no competing financial or non-financial interests.